\documentstyle[prl,twocolumn,aps]{revtex}
\begin{document}
\draft
\twocolumn[\hsize\textwidth\columnwidth\hsize\csname @twocolumnfalse\endcsname

\title{Pseudoparticle Description of the 1D Hubbard Model \\
Electronic Transport Properties} 

\author{N. M. R. Peres$^{(1,2)}$, J. M. P. Carmelo$^{(1)}$,
D. K. Campbell$^{(2)}$, and A. W. Sandvik$^{(2)}$}

\address{$^{(1)}$ Departamento de F\'{\i}sica, Universidade de \'Evora,
Apartado 94, P-7001 \'Evora Codex, Portugal}
\address{$^{(2)}$ Department of Physics, University of Illinois at
Urbana-Champaign, 1110 West Green Street, Urbana, Illinois 61801}

\date{\today}

\maketitle

\begin{abstract}
We extend the pseudoparticle transport description of the
Hubbard chain to all energy scales. In particular we compute the
mean value of the electric current transported by any Bethe-ansatz
state and the transport masses of the charge carriers.
We present numerical results  for the optical conductivity of the model
at half-filling for values of $U/t=3$ and $4$. We show that these are
in good agreement with the pseudoparticle description of the 
finite-energy transitions involving new pseudoparticle energy bands.
\end{abstract}
\vskip2mm
\vskip2mm]

\section{Introduction}
The electronic transport properties of strongly correlated electron systems
in low-dimensional conductors have been
of experimental and theoretical interest for a long time.
Low dimensional conductors show
large deviations from the usual-single particle description which suggests
that electronic correlations might play an important role \cite{Jacobsen}.
Although the Hubbard chain has been diagonalized by Lieb and Wu
\cite{Lieb} long ago, the involved form of the Bethe ansatz (BA)
wave function has prevented the calculation of dynamic response functions,
like
the optical conductivity $\sigma(\omega)$. Several approaches,
using perturbation theory \cite{Maldague}, the pseudoparticle
formalism \cite{Carm4}, bosonization \cite{Gimarchi},
scaling methods \cite{Stafford}, and spin wave theory
\cite{Horsch} have been used to investigate the low energy behavior of
$\sigma(\omega)$ away from half-filling and at the metal-insulator
transition \cite{Lieb}. Interesting information on
of $\sigma(\omega)$ at finite values $\omega$ has been
obtained numerically \cite{Loh,Fye,Sandvik}.

Recently, it has been possible to obtain analytical results for the
finite-$\omega$ behavior of $\sigma(\omega)$ using a pseudoparticle
description for all the Hamiltonian eigenstates of the model \cite{Nuno1}.
This theory generalizes, for all energy scales, previous results
by Carmelo {\it et al.} \cite{David}, by introducing new branches
of pseudoparticles. The latter are called heavy-pseudoparticles
because the eigenstates described by them
have an energy gap relatively to the ground state.
As in the case of the low-energy properties of the 1D Hubbard model
\cite{David}, it is possible to write the Hamiltonian in terms of a set of
anti-commuting operators, called pseudoparticle operators, and to generate
all the model eigenstates from the $SO(4)$ ground state \cite{Nuno3}.
We show below that the finite-$\omega$ properties of the optical
conductivity at half filling are determined by transitions
involving the bands populated by the usual
pseudoparticles (called $c$ and $s$ pseudoparticles \cite{David})
and the heavy-pseudoparticle bands (called $c,\gamma$ and $s,\gamma$
pseudoparticles \cite{Nuno1}).

The paper is organized as follows. In Section II we study, within the
pseudoparticle formalism, the response of
the Hubbard ring to a small electric field and compute the mean
value of the electronic current for any BA state and the
transport masses of the charge carriers. In Section III we present numerical
results for the optical conductivity of the half-filled
1D Hubbard model and compare them with the pseudoparticle-theory
predictions. In Section IV we summarize our results and discuss
possible future developments.
\section{Response to an Electric Field: Pseudoparticle Description}

The linear response of the 1D Hubbard model to a small spatially
uniform, time-dependent, electric field parallel to the chain
is achieved by means of the usual Peierls phase substitution of the hopping
integral ($t \rightarrow te^{\pm i\Phi(t)/N_a}$). This means that one threads
the ring by a flux $\Phi(t)$ \cite{Maldague,Shastry}. The obtained
Hubbard-chain Hamiltonian (with $N$ electrons) in a magnetic field $H$
and with chemical potential $\mu$ is written as

\begin{eqnarray}
        \hat{H} &=& -t\sum_{j,\sigma}
        \left[e^{+i\Phi(t)/N_a}c_{j\sigma}^{\dag}c_{j+1\sigma} +
        h. c. \right] \nonumber\\
        &+&U\sum_{j} \hat{D}_j
        -\mu(N_a-\sum_{j,\sigma}\hat{n}_{j,\sigma})
        -\mu_0 H \sum_{\sigma,j}\sigma\hat{n}_{j,\sigma}\, ,
\label{hamilt}
\end{eqnarray}
where $\hat{D}_j=[\hat{n}_{j,\uparrow}-1/2][\hat{n}_{j,\downarrow}-1/2]$,
$c_{j\sigma}^{\dag }(c_{j\sigma})$ creates (annihilates)
an electron with spin
$\sigma$ ($\sigma = \uparrow,\downarrow$ when used as operator index or
$\sigma=\pm1$ otherwise),
$\hat{n}_{j,\sigma}=c_{j\sigma}^{\dag}c_{j\sigma}$ is the number
operator at site $j$, $N_a$ is the number of sites of the ring, and
$c_{N_a+1\sigma}=c_{1\sigma}$ (we are using units such that $\hbar=1$,
the lattice spacing $a=1$, and the electron charge $e=-1$).
The form of the interaction term accounts for the
particle-hole symmetry of the model at half filling \cite {Lieb,Bill}.

Although  the pseudoparticle description refers to all Hamiltonian
eigenstates \cite{Nuno1}, in this paper we restrict our study to the
Hilbert subspace involved in the zero-temperature conductivity spectrum.
Since the current operator commutes with the eta-spin and spin operators
$\hat{\eta}^2$, $\hat{\eta}_z$ and $\hat{S}^2$, $\hat{S}_z$, respectively,
the Hilbert subspace is in the present parameter space spanned by the
lowest-weight states (LWS's) of the eta-spin and spin algebras
\cite{Yang}.

It has been possible to solve the Hamiltonian (\ref{hamilt}) by means of the
coordinate BA both with twisted and toroidal boundary conditions
(both approaches give essentially the same results)
\cite{Shastry,Martins}. One obtains the energy spectrum of the model parameterized
by a set a numbers $\{k_j,S_{\gamma,j},R_{\gamma,j}\}$
which are solution of the BA interaction equations.
Following Takahashi's $\Phi(t)=0$ formalism \cite{Takahashi}, the interaction
equations with the flux $\Phi(t)$ read

\begin{eqnarray}
        k_jN_a&=&
        2\pi I_j^c+\Phi(t)-\sum_{\gamma=0}^\infty \sum_{j'=1}^{N_{s,\gamma}}
        \theta\left(\frac{sin(k_j)/u-S_{\gamma,j'}}{(\gamma+1)} \right)-
        \nonumber\\
        &-&\sum_{\gamma=1}^\infty \sum_{j'=1}^{N_{c,\gamma}}
        \theta\left(\frac{sin(k_j)/u-R_{\gamma,j'}}{\gamma}\right)\, ,
\label{tak1}
\end{eqnarray}
\begin{eqnarray}
        2N_a &R&e \, sin^{-1}[(R_{\gamma,j}+i\gamma)u]=2\pi I^{c,\gamma}_j
        + 2\gamma \Phi(t)-\nonumber\\
        &-&\sum_{j'=1}^{N_c}\theta\left(\frac{sin(k_{j'})/u-R_{\gamma,j}}
        {\gamma}\right)+\nonumber\\
        &+&\sum_{\gamma'=1}^\infty \sum_{j'=1}^{N_{c,\gamma'}}
        \Theta_{\gamma,\gamma'}(R_{\gamma,j}-R_{\gamma',j'})\, ,
\label{tak2}
\end{eqnarray}
and
\begin{eqnarray} 
        2\pi I^{s,\gamma}_j&=&
        \sum_{j'=1}^{N_c}\theta\left(\frac{S_{\gamma,j}-sin(k_j')/u}
        {(\gamma+1)}\right)-\nonumber\\
        &-& \sum_{\gamma'=0}^\infty \sum_{j'=1}^{N_{s,\gamma'}}
        \Theta_{\gamma+1,\gamma'+1}(S_{\gamma,j}-S_{\gamma',j'})\, .
\label{tak3}
\end{eqnarray}
In Eqs. (\ref{tak1}), (\ref{tak2}), and (\ref{tak3}) we have that
$\theta=2tan^{-1}(x)$, $u=U/4t$, and the functions 
$\Theta_{\gamma,\gamma'}$, and
$\Theta_{\gamma+1,\gamma'+1}$ have the same definition
as in Takahashi's paper \cite{Takahashi}. The following definitions,
$\Lambda_\alpha^{n+1}/u=S_{\gamma,j}$ (with $n+1=\gamma$ and $\alpha=j$),
$\Lambda_\alpha^{' \, n}/u=R_{\gamma,j}$ (with $n=\gamma$ and $\alpha=j$)
with $\gamma=1,2,\ldots,\infty$ for the $N_{c,\gamma}$ sums and
$\gamma=0,1,2,\ldots,\infty$ for the $N_{s,\gamma}$ sums,
allows us to recover Takahashi's formulae for $\Phi(t)=0$ \cite{Takahashi}
(often we also use the notation $c\equiv c,0$, with the $c,\gamma$
sums running over $0,1,2,\ldots,\infty$).
The quantum numbers $I_j^c$, $I^{c,\gamma}_j$, and $I^{s,\gamma}_j$
are consecutive integers or half-odd integers depending on the
parity of the numbers $\sum_{\gamma=0}N_{s,\gamma}+\sum_{\gamma=1}
N_{c,\gamma}$, $N_a-N+N_{c,\gamma}$, and $N-N_{s,\gamma}$, respectively.
The conservation of the electrons numbers imposes the sum rules
$N_\downarrow=\sum_{\gamma=1}\gamma
N_{c,\gamma}+\sum_{\gamma=0}(\gamma+1)N_{s,\gamma}$
and $N_c=N-2\sum_{\gamma=1}\gamma N{c,\gamma}$ on the numbers $N_c$,
$N_{c,\gamma}$, and $N_{s,\gamma}$. All the LWS's
of the model \cite{Yang} are determined
by different occupancies of these quantum numbers. For example,
the ground state is described by a compact symmetric occupancy around the
origin of the numbers $I_j^c$ and $I_j^{s,0}$, and by zero occupancy for
the numbers $I^{c,\gamma}_j$ and $I^{s,\gamma>0}_j$ \cite{Lieb}.
Equivalently, the eigenstates of the model can 
be described in terms of pseudomomentum
$\{q_j^{\alpha,\gamma}=2\pi I_j^{\alpha,\gamma}/N_a\}$ distributions,
where $\alpha=c,s$ and $I_j^{c,0} \equiv I_j^c$.

The energy eigenvalues are given by

\begin{eqnarray}
        E&=&-2t\sum_{j=1}^{N_c}cos(k_j)+
        \sum_{\gamma=1}^\infty \sum_{j=1}^{N_{c,\gamma}}4t Re
        \sqrt{1-u^2(R_{\gamma,j}+i\gamma)^2}\nonumber\\
        &+&N_a(U/4-\mu)+N(\mu-U/2)-\mu_0 H(N_\uparrow-N_\downarrow)\,.
\label{energy}
\end{eqnarray}
In the limit of a large system ($N_a\rightarrow \infty$, $N/N_a$ fixed)
we can develop a generalization for the low-energy Landau-liquid description
of the 1D Hubbard model \cite{Carm1,Carm2,Carm3} and of its
operational representation \cite{David} for all energy scales
\cite{Nuno1}. Within this new general framework we have the usual
$c,0\equiv c$ and $s,0\equiv s$ pseudoparticles (and their energy bands),
whose different occupancies of the pseudomomenta
$\{2\pi I_j^{c,0}/N_a\}$ and $\{2\pi I_j^{s,0}/N_a\}$, respectively,
describe the gapless low-energy physics. In addition, we have the branches
(and energy bands) of heavy $c,\gamma$ and $s,\gamma$ pseudoparticles, whose
different occupancies of the pseudomomenta
$\{2\pi I_j^{c,\gamma>0}/N_a\}$ and $\{2\pi I_j^{s,\gamma>0}/N_a\}$,
respectively, describe excited eigenstates with an energy gap relatively
to the ground state \cite{Nuno1}.

Combining Eqs. (\ref{tak1}-\ref{energy})
there are several interesting transport quantities that can be computed.
An example of physical interest for the study of metal-insulator transitions
is the charge stiffness $D$ 
\cite{Carm4,Stafford,Fye,Shastry,Millis,Scalapino}.
Information on $D$ is at zero temperature
contained in the ground state
energy $E_0$ through the relation \cite {Kohn}

\begin{equation}
        D=\left. \frac 1{2}
        \frac{d^2(E_0/N_a)}{d(\Phi(t)/N_a)^2}\right\vert_{\Phi(t)=0} \, .
\label{drude}
\end{equation}
This quantity can be computed within the pseudoparticle formalism by
combining a Boltzmann transport description for the pseudoparticles
with linear response theory \cite{Carm4} or by direct evaluation of
Eq. (\ref{drude}) \cite{Stafford,Nuno2}. Another important transport 
quantity is the electric
current of a given eigenstate, with the operator $\hat{j}$ reading,
for the Hubbard model, $\hat{j}=it\sum_{j=1}^{N_a}(c_{j\sigma}^{\dag}c_{j+1\sigma}-c_{j+1\sigma}^
{\dag }c_{j\sigma})$.
Since in this model $\hat{H}$ and $\hat{j}$ do not commute
the BA wave function does not diagonalizes
simultaneously the Hamiltonian (\ref{hamilt}) and the current operator
$\hat{j}$. Therefore, the BA alone can only provide 
information on the mean value of the electric current.
This is an important quantity for it allows us to
compute the transport masses of the charge carriers of the system. For the
Hubbard chain, the carriers are the pseudoparticles $c$ and
$c,\gamma$, as it follows from Eqs. (\ref{tak1}-\ref{tak3}) and from
a alternative
Boltzmann transport analysis \cite{Nuno2}. The coupling constants to the
electric field are given by ${\cal C}^{\rho}_c=1$
and ${\cal C}^{\rho}_{c,\gamma}=2\gamma$ (and ${\cal C}^{\rho}_{s,\gamma}=0$).

The mean value of the current operator $<m\vert \, \hat{j}\vert m>$
transported by the eigenstate $\vert m>$ of energy $E_m$ is given by
$-{d(E_m/N_a)}/{d(\Phi(t)/N_a)}\vert_{\Phi(t)=0}$. As stated above,
the LWS's are obtained considering all the possible occupation
distributions of the pseudomomenta $2\pi I_j^{\alpha,\gamma}/N_a$. 
For this reason, it is simpler to describe
$<m\vert \, \hat{j}\vert m>$ in terms of the pseudoparticle occupation
distributions $N_{\alpha,\gamma}(q)$ ($q$ being the pseudomomentum).
The computation of $<m\vert \, \hat{j}\vert m>$ involves 
the expansion of Eqs.
(\ref{tak1}-\ref{energy}) up to first order in the flux
$\Phi(t)$. After lengthy calculations, we obtain (in the limit of $N_a\rightarrow\infty$)

\begin{equation}
        <m\vert \, \hat{j}\vert m>=\sum_{\alpha=c,s}
        \sum_{\gamma=0}^{\infty}
        \int_{-q_{\alpha,\gamma}}^{q_{\alpha,\gamma}}dq
        N_{\alpha,\gamma}(q)
        j_{\alpha,\gamma}(q)\, ,
\label{deltaj}
\end{equation}
where $j_{\alpha,\gamma}(q)=\sum_{\alpha'=c,s}\sum_{\gamma'=0}^{\infty}
{\cal C}^{\rho}_{\alpha',\gamma'}[v_{\alpha,\gamma}(q)\delta_{\alpha ,\alpha'}
\delta_{\gamma ,\gamma'}+{1\over 2\pi}\sum_{j=\pm1}jf_{\alpha,\gamma;\alpha'
,\gamma'}(q,jq_{F\alpha',\gamma'})]$, and $v_{\alpha,\gamma}(q)$ are the pseudoparticle velocities
(defined as $d\varepsilon^0_{\alpha,\gamma}(q)/dq$, where
$\varepsilon^0_{\alpha,\gamma}(q)$
is the pseudoparticle dispersion), $f_{\alpha,\gamma;\alpha',\gamma'}(q,q')$ are the pseudoparticle interactions
(or $f$-functions), and
the momenta $q_{F\alpha,\gamma}$ are the pseudo-Fermi points  \cite{Nuno1}.
The first term
of $j_{\alpha,\gamma}(q)$ is what we would expect for a non interacting gas
of pseudoparticles and the second term describes 
the dragging effect on a single 
pseudoparticle due to its interactions with all the others. 

We stress that
Eq. (\ref{deltaj}) is valid for all energies because, in contrast to
the Fermi liquid theory, there is only forward scattering among the
pseudoparticles at all energy scales. The charge transport
masses $m_{\alpha,\gamma}^{\rho}$ are defined as $m^{\rho}_{\alpha,\gamma}=
q_{F\alpha,\gamma}/ [{\cal C}^{\rho}_{\alpha,\gamma}j_{\alpha,\gamma}(q)]$
and contain important physical information. An interesting and well known
property is the metal-insulator transition of the model which occurs
at half-filling and zero temperature \cite{Lieb}.
It has been shown by means of the pseudoparticle formalism
\cite{Carm4} that, at half-filling, the transport mass of the $c$
pseudoparticles is infinite (whereas $m_{s,\gamma}^{\rho}=\infty$
always), as it
should be for an insulator, and that the charge stiffness is also zero 
($D\propto1/m_c^\rho$ \cite{Carm4}), satisfying Kohn criterion \cite{Kohn}.

In contrast to the zero temperature limit, the $c,\gamma$
heavy pseudoparticles
play an important role in the transport properties of the model. We
computed the transport masses for the $c,\gamma$ heavy pseudoparticles 
in the limit of half filling (in a zero magnetic field) and obtained
$m_{c,\gamma}^{\rho} \rightarrow \infty$ as $n \rightarrow 1$.
Whether this result
can shed some light on the relation between integrability and anomalous
transport properties \cite{Castella1} is something that 
requires future studies.

\section{Optical Conductivity at Half-Filling}

As discussed in the previous Section, some transport properties of the
model can be understood from the use of the general
pseudoparticle theory. At half filling and zero temperature
the charge stiffness is zero and the real part of the optical conductivity
$Re \sigma (\omega)$ has a finite
energy gap for all values of $U/t$. At half filling the $c,0$
pseudoparticle band is full and excitations can only occur between
the low-energy $\alpha$ bands and the heavy-pseudoparticle bands. Thus,
the form of the optical conductivity is determined by these type of
excitations, which are rather unusual. For example, to populate the
lowest $c,\gamma$ energy-band (the $c,1$ band) with one $c,1$ heavy
pseudoparticle one has to annihilate
two $c$ and one $s$ pseudoparticles. The energy involved in this process
is $\epsilon^0_{c,1}(0)-\epsilon^0_{c,0}(q_{Fc,0})-
\epsilon^0_{c,0}(-q_{Fc,0})-\epsilon^0_{s,0}(q_{Fs,0}) =2\mu(n=1)$,
where $\mu(n=1)$ stands for the chemical potential in the limit of
half filling which equals the Mott-Hubbard gap \cite{Nuno1}.
In addition, this excitation process is characterized by
a topological excitation of the $s$ pseudoparticles of momentum
$-q_{Fs,0}$ \cite{Nuno3}, ensuring that the overall momentum
vanishes, as required for the excitations induced by the
current operator. In general,
the process of creating one $c,\gamma$ heavy pseudoparticle has a minimum
energy cost of $2 \gamma \mu(1)$ and a maximum cost of $2 \gamma \mu(1)+
8 \gamma t$. These transitions are expected to show up in
the optical conductivity as a set a optical bands.

In order to investigate this problem, we have studied the model
numerically\cite{Sandvik}, by using a quantum Monte Carlo technique
for systems
with 32 sites\cite{Qmc}, which is  considerably larger than 
previous exact diagonalization studies\cite{Loh,Fye}. The current-current
correlation functions are computed in imaginary time and continued to real
frequency using the so called maximum-entropy method\cite{Maxent}. The
resolution of this method is limited but, nevertheless, gives a
semi-quantitative confirmation of a multi-band optical conductivity.

In
Fig.~\ref{fig1} we show our results for $U/t=3$ and $4$ where two peaks are
clearly resolved. As indicated in the figure, the lower edges of the spectra
agree very well with the predicted gaps, and the second peak appears above
the predicted lower edge of the second band, as obtained from the
pseudoparticle theory, which also allows the evaluation of the 
critical exponents associated with the conductivity edges \cite{Nuno1}.
Note that the weight of the first band
is concentrated towards the lower end of the allowed band of width $8t$.
The second peak seen in our results is probably dominated by the second
optical band, but likely contains contributions also from the tail of the
first band (higher bands cannot be resolved due to the
limitations of the method).

As
discussed in detail in Ref.~\cite{Sandvik}, for large values of
the Hubbard interaction $U$ only a single broad peak of width $\approx 8t$
is observed, likely due to the low intensity of the higher bands.


\section{Summary and Open Questions}

In this paper we developed a general theory, valid for all energy
scales, for the electronic transport properties of the Hubbard 
chain (details of the calculations will be presented elsewhere \cite{Nuno2}).
The fact that
$[\hat{H},\hat{j}]$ is non zero does not allow us to obtain information on
the current operator $\hat{j}$ it self from the BA equations but only on
its mean value.
(The conductivity exponents can be obtained by combining BA and
properties following from the model infinite conservation laws.)
Still, this is an important quantity, since it is related
to the transport mass of the charge carriers (the $c,\gamma$
pseudoparticles, with $\gamma=0,1,\ldots,\infty)$ which is believed to
be important in the transport properties of the model at finite temperatures. This question is now an issue of ongoing research.

In addition, using maximum-entropy analytic continuation of quantum
Monte Carlo data, we computed the optical conductivity of $32$-site Hubbard
chains for $U/t=3$ and $4$, and found a two-band structure with band edges
at energies consistent with our pseudoparticle theory. This agrees with the
conductivity absorption edges obtained directly from
the pseudoparticle theory.


\begin{figure}
\caption{Optical conductivity results for $32$-site Hubbard chains
at half-filling, for values of $U/t=3$ (top) and $4$ (bottom), computed
using maximum-entropy analytic continuation of quantum Monte Carlo data.
The solid and dashed vertical lines indicate the theoretical predictions
for the lower edges of the first two bands.}
\label{fig1}
\end{figure}
\end{document}